# Wide but Shallow: Coding Agents in Generative Art

Dominikus Baur[1]

[1] University of Applied Sciences Munich (HM), Munich, Germany
dominikus.baur@hm.edu

**Abstract.** This paper presents an autoethnographic study on the use of coding agents in code-based generative art. Over the course of the 31-day *#Genuary* art challenge, the author examined and compared a manual approach with the use of coding agents in their practice. The growth in quality of coding agents enables their use to autonomously produce programming code. Generative art's bricolage quality of combining existing algorithms and a coding agent's ability to draw on extended training data leads to a difficulty collapse: nearly any well-known algorithm can be used directly. While this enables easier free-form experimentation and a wider variety of outputs, it comes at the cost of less depth. Overall, this points towards a slow progression in artists' roles from inventing algorithms to changing parameters of existing ones and a potential overall artistic surrender to the results from large-language models.

## 1 Introduction

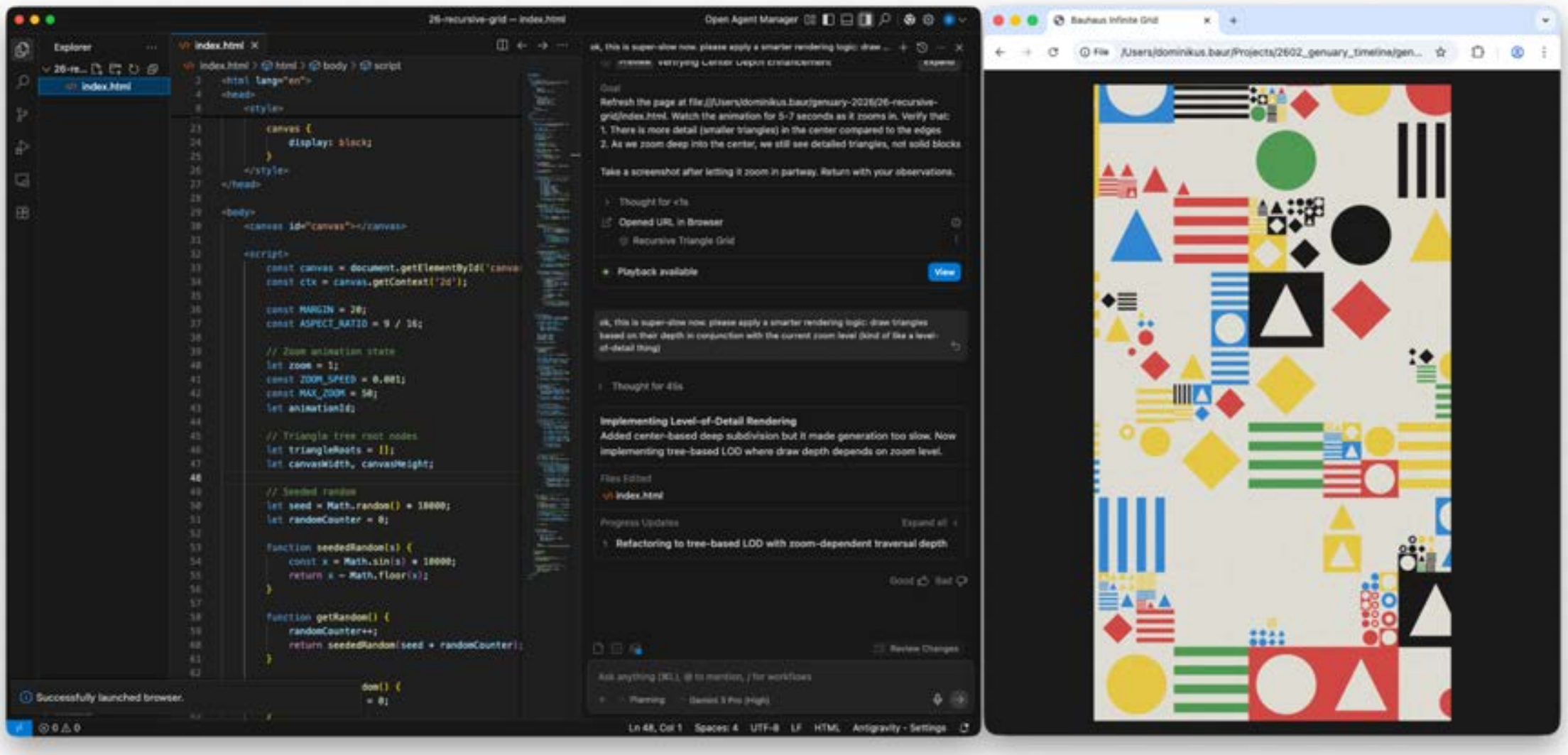

**Fig. 1** - Coding-agent supported generative art in Google's Antigravity IDE.

The level of quality large-language models (LLMs) have reached especially for coding situations (Anthropic 2025) enables a new trend: *vibe coding* (Karpathy 2025). Programming small tools and websites(Karpathy 2025) without any programming experience or interference with the code is spreading within communities of amateurs (Edwards 2026; Fawzy et al. 2025). Additionally, professional software developers also increasingly rely on these tools (Cui et al. 2025). Extensive tool support and integrations with integrated development environments (IDEs) (e.g., Google Antigravity, see **Fig. 1**) have turned the underlying LLMs into coding agents.

Writing code has also always been a central part of code-based or algorithmic generative visual art (Galanter 2003) (not to be confused with text-to-image generative AI art (Grba 2022; McCormack et al. 2023)). Code-based generative art has the artists (sometimes self-described as Algorists (Verostko 2021)) create autonomous systems - usually in computer code - that produce the resulting artworks, thus creating an indirection between artist and artwork, funnelling the artist's intent through the code. Using programmed systems let artists draw on a variety of algorithms from physics, biology, mathematics and computer graphics (cf. Shiffman 2024). Also, generative artworks rarely exhibit a fixed form: integrating mechanisms for randomness usually means that every run of the algorithm produces a novel output within an intentional design space defined by the artist through the algorithm.

It is to be expected that the vibe coding trend and coding agents will also have an impact on a practice as coding centric as generative art. The question is: how will this change the practice and its practitioners?

As a generative artist and researcher[1], I decided to investigate the impact the use of a coding agent has on my own practice. To this end, I've used an autoethnographic approach (Ellis et al. 2011) to study the effect as well as derive generalizable results. Adding to initial experiments by generative artists with coding agents (e.g., Goodchild 2023) and interviews with educators (McNutt et al. 2025) and machine learning-oriented artists (Rozental and de Rooij 2026), this work contributes the "interior" perspective through the autoethnographic approach.

For my methodology, I took the month-long *#Genuary 2026* art challenge[2] where artists are expected to produce daily artworks based on predefined artistic prompts. I've created 31 generative artworks[3] and split up my approach into three parts: starting with pure manual coding (4 days), switching over to loose coding agent support (11 days) to gather an initial understanding of the effects, and finally more detailed notetaking for the remaining 16 days of the challenge.

The qualitative coding of the resulting notes shows the impact the support from coding agents has on the practice, leading to a difficulty collapse where seemingly every algorithm - no matter how complex - becomes usable to the artist thus enabling more variety and experimentation across their work. This, however, comes at the cost of making outputs shallower due to the lack of understanding and thus the inability to fine-tune them. The study also points at a slow progression towards more parametric than coding interaction by artists and the potential risk of an overall artistic surrender to the decisions of the machine.

# 2 Generative Art and Coding Agents

One of the standard definitions of generative art by Galanter emphasizes the ceding of control by the artist to an autonomous system:

> Generative art refers to any art practice where the artist uses a system, such as a set of natural language rules, a computer program, a machine, or other procedural invention, which is set into motion with some degree of autonomy contributing to or resulting in a completed work of art. (Galanter 2003)

While these systems are sometimes conceived from the ground up by an artist, more commonly they are based on or inspired by *algorithms* from various fields that describe the behavior of complex systems through computation: complexity theory and mathematics (fractals, Lindenmayer systems, cellular automata), biology (evolution, ant-, flocking-, and slime mold simulations), chemistry (reaction-diffusion systems), physics (physics simulations, pendulums), etc. (Galanter 2016; Shiffman 2024). Other commonly used algorithms are noise functions to turn randomness into natural-looking results (Perlin 2002) and techniques from computer graphics such as path tracing (Kajiya 1986) for rendering 3D scenes. Grba compares this adoption of "tools, materials, and artefacts available from the immediate surroundings" with the bricolage techniques of the 1960s (Grba 2023). The artist's contribution is thus the combination and adjustment of these various existing parts.

---

[1] For my artistic work see https://dominikus.art
[2] https://genuary.art
[3] See the results, models used, chatlogs, etc. at https://dominikus.art/genuary2026

Generative artists hereby also embrace the "open source" idea of all science: the algorithms used are often re-implemented from the corresponding research papers and shared online as help for others. As Galanter writes: "[t]his type of creative approach breaks with the paradigm of the heroic single artist creating a 'fixed' masterpiece. It creates a process where multiple artists evolve an ever changing and growing family of related artworks over time" (Galanter 2016).

Additionally, generative artists often take pride in the code they write, since they exhibit direct control only over this, not the resulting visual pieces. Artists like Piter Pasma[4] and others, for example, engage in "code golf"[5], a practice where code that produces the same visual output should be as short and concise as possible, with a range of tricks and processes available to arrive at this.

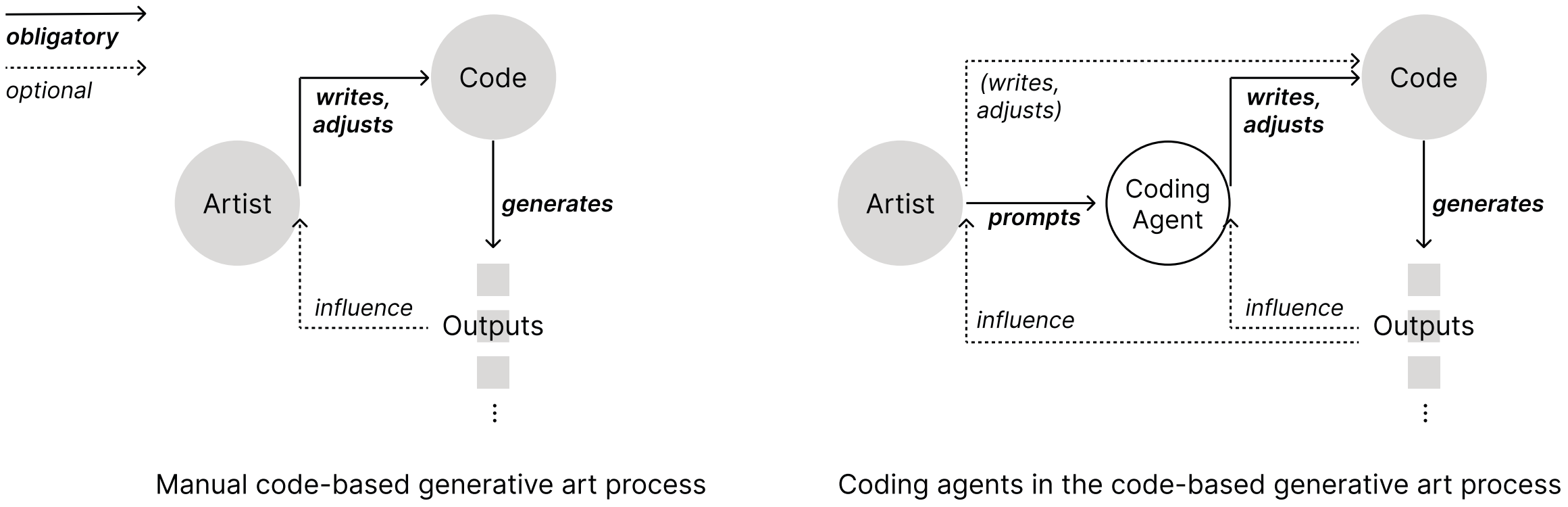

**Fig. 2** - The code-based generative art process, both manual and including coding agents. Note the cyclical nature and the similarity to the generative *design* process (cf. Bohnacker et al. 2012).

## 2.1 The generative art process

In the code-based, visual variety of generative art, artists write computer programs that generate static, animated and/or interactive outputs. Since the first version of the code might not yield the intended visual output, artists usually enter a cycle of comparing the outputs to their artistic vision, adjusting the code, comparing again and so on (see **Fig. 2** (left)).

Here is how generative artist Vera Molnár describes her process with the IBM 370 computer in 1975:

> When I have an idea for a new picture, I make the first version of it rather quickly. Usually I am more or less dissatisfied with it and I modify it. I alter in a stepwise manner the dimensions, proportions and arrangement of the shapes. When simple geometrical shapes are used, such modifications are relatively easy to make. By comparing the successive

[4] https://piterpasma.nl
[5] https://codegolf.stackexchange.com

> pictures resulting from a series of modifications, I can decide whether the trend is toward the result that I desire. (Molnár 1975)

This cyclical process is core to the practice and corresponds with a shrinking gap between artistic intent and output (see **Fig. 3**). Artists start with a rather high-level description of their intent or idea: they write complete program code that contains the basic algorithm(s) that they want to use (whether existing or newly made-up ones). Looking at the outputs, they then dive deeper, adjusting parts of the algorithms or functions. Once the then-created outputs are more satisfactory, they start changing parameters and finally maybe even single numbers at the lowest level of the code to arrive at the final version of code and artwork (cf. Hobbs 2014; Verano Merino and Sáenz 2023).

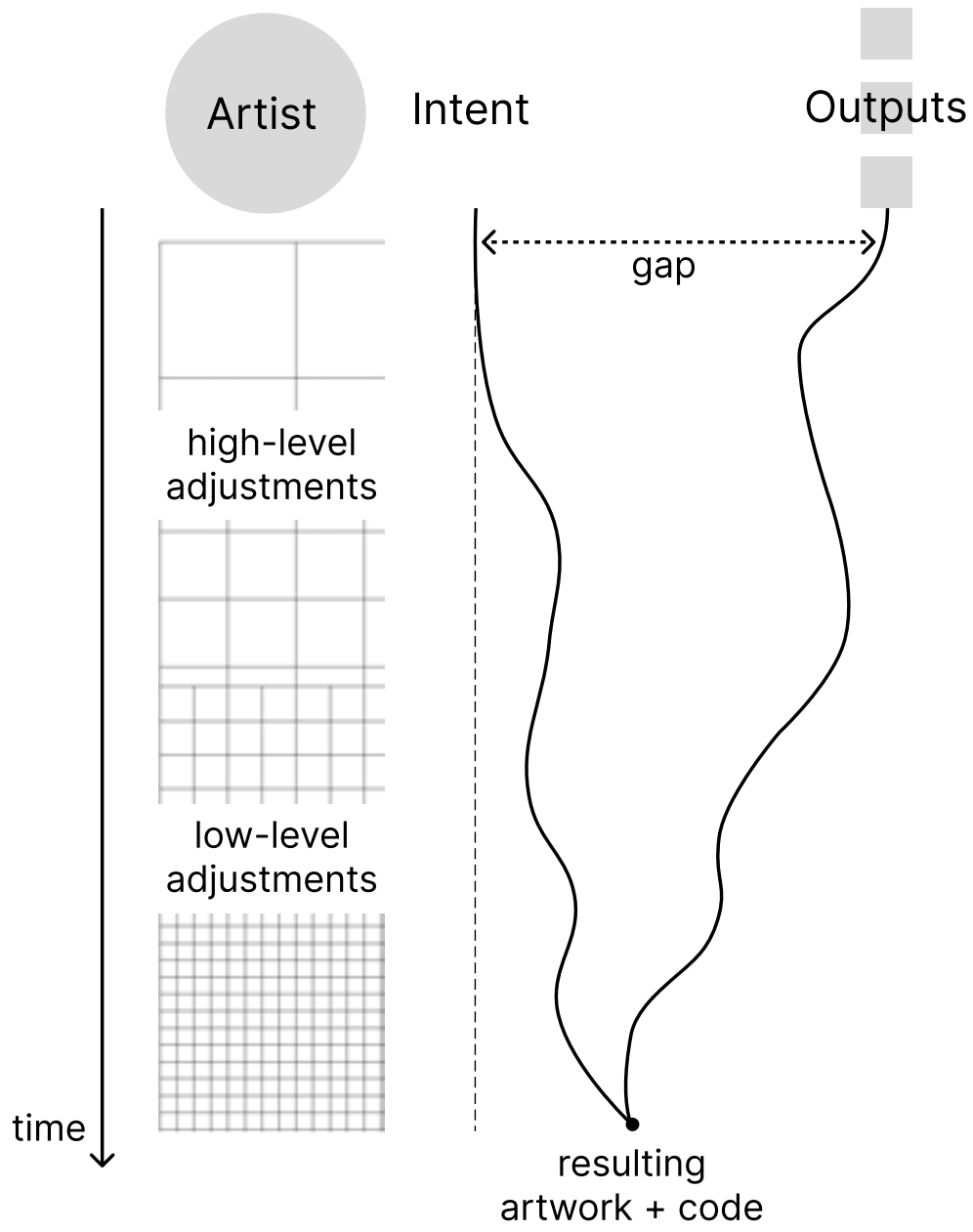

**Fig. 3** - The generative art process as shrinking the gap between artistic intent and output.

Several aspects thus coincide in this (idealized) version of the process over time which sees the artists descend a ladder of abstraction in their code (**Fig. 3**): the gap between artistic intent and output shrinks. Additionally, artists move from high-level (broad strokes) to low-level adjustments (specific and detailed) in their code. Thus, they're also gaining more control over the actual results, specifying it in greater detail over time. Note also that the artistic intent is not fixed here: aspects of the output might shift it towards or away from its original state, making the outcome always a result of compromise.

What complicates matters is the resistance and quality of the material in the generative art practice. To put it with Malafouris' metaphor (Malafouris 2008): just like a potter interacts with the resisting clay thus creating a cycle feeding on both artistic and material agency, the generative artist tries to impose their ideas onto code, a material with its own specific qualities and resistances.

What gives code-based generative art an especially challenging quality as a material is its reliance on systemic emergence. In generative art this refers to setting up a system that produces unexpected and complex outputs (McCormack and Dorin 2001). The nature of such systems usually prevents specific predictions what the resulting output might be once the computation has completed. With computational systems, changing even single lines of code can completely alter a result. Therefore, the artists' imaginations of what their system would produce are often undermined by the reality of the output. As artist Marius Watz says:

> The artist specifies the initial boundaries and strategies of creation, and then enters into a feedback loop of adjusting parameters in a search for optimal regions in parameter space. The moment of genuine surprise is often the moment of breakthrough. (Watz 2006)

## 2.2 Coding Agents

Introducing LLM-based coding agents into this process creates an additional layer of indirection (see **Fig. 2** (right)): instead of writing the computer programs themselves, artists instead prompt the coding agent to produce the code for them. Said code then produces the artistic output, just like in the original process.

Code generation via large-language models has gone through significant changes over the last years: starting from generating snippets of code in a chatbot window, via code completion within IDEs (GitHub Copilot[6]), the state-of-the-art is now coding agents that can parse entire codebases and use tools like web browsers and file systems (Dong et al. 2025).

An interesting attribute of coding agents is that they are very good at replicating code that they have encountered repeatedly in their training data (e.g., Katzy et al. 2024) – no matter the complexity of that code. Therefore, we have an effect that I came to call *difficulty collapse*: while people need more time to write complex than simple code, for the machine only the length of the code to be generated matters. One hundred lines of intricate and complex shader code takes (basically) the same amount of time to generate than one hundred lines of meaningless numbers. Therefore, to get highly complex code from a coding agent, all one needs is the right term (i.e., the name of the algorithm) in the prompt.

One could argue that this would make them a fit for generative artists and the bricolage atmosphere of the generative art scene: asking a coding agent for an implementation of a reaction-diffusion system (Gray and Scott 1983) usually directly leads to working code with the artist then being able to continue the process on this foundation. Manually re-implementing the already existing algorithm would lead to roughly the same result, but with a larger time investment.

Also, generative artists are used to handing over control of their art to a system. The indirection is part of the practice. Therefore, it seems like this modification of the process keeps the central gradient of lessening abstraction (**Fig. 3**) intact: instead of starting at programming an algorithm, the artist begins at prompting an agent in natural language and/or pseudocode, thus introducing merely a new step in the abstraction ladder.

Yet coding agents go beyond that: they can interfere and support at any point in the process, from switching over to a completely new algorithm to tweaking single numbers in the code. Integrating them goes beyond using a specific tool like a pen plotter, for example, which - while enabling certain outputs and a certain look-and-feel - leaves the creative process altogether with the artist. An agent, however, can take over certain aspects completely and thus act more like an apprentice in this metaphor that makes independent artistic choices.

And this use of autonomy might not even be intentional on the artist's side: prompting is naturally less precise than writing code (McCormack et al. 2023), which means that the coding agent has to make many implementation decisions autonomously. Additionally, the artist's intent for their artwork is not fixed - as we've seen - but can be swayed by the quality of the outputs and/or the time available. Thus, autonomous decisions by the agent could over time accumulate to move the resulting artwork even further away from the artist's original intent than in a manual process (see **Fig. 4**).

[6] https://github.com/features/copilot

As depicted in **Fig. 2** (right), the coding agent indeed takes over the place of the artist from the manual generative art process: IDEs like Google's Antigravity[7] (see **Fig. 1**) let the agent even perform the evaluation step by enabling it to not only generate program code, but also deploy it to a web browser, make screenshots and screen recordings, comparing them with the original requirements from the prompt and adjusting the code if necessary. If it wasn't for the lack of an initial prompt, the agent could act completely autonomously, thus removing the artist from the process altogether, something that is already happening in regular software development (Li et al. 2025).

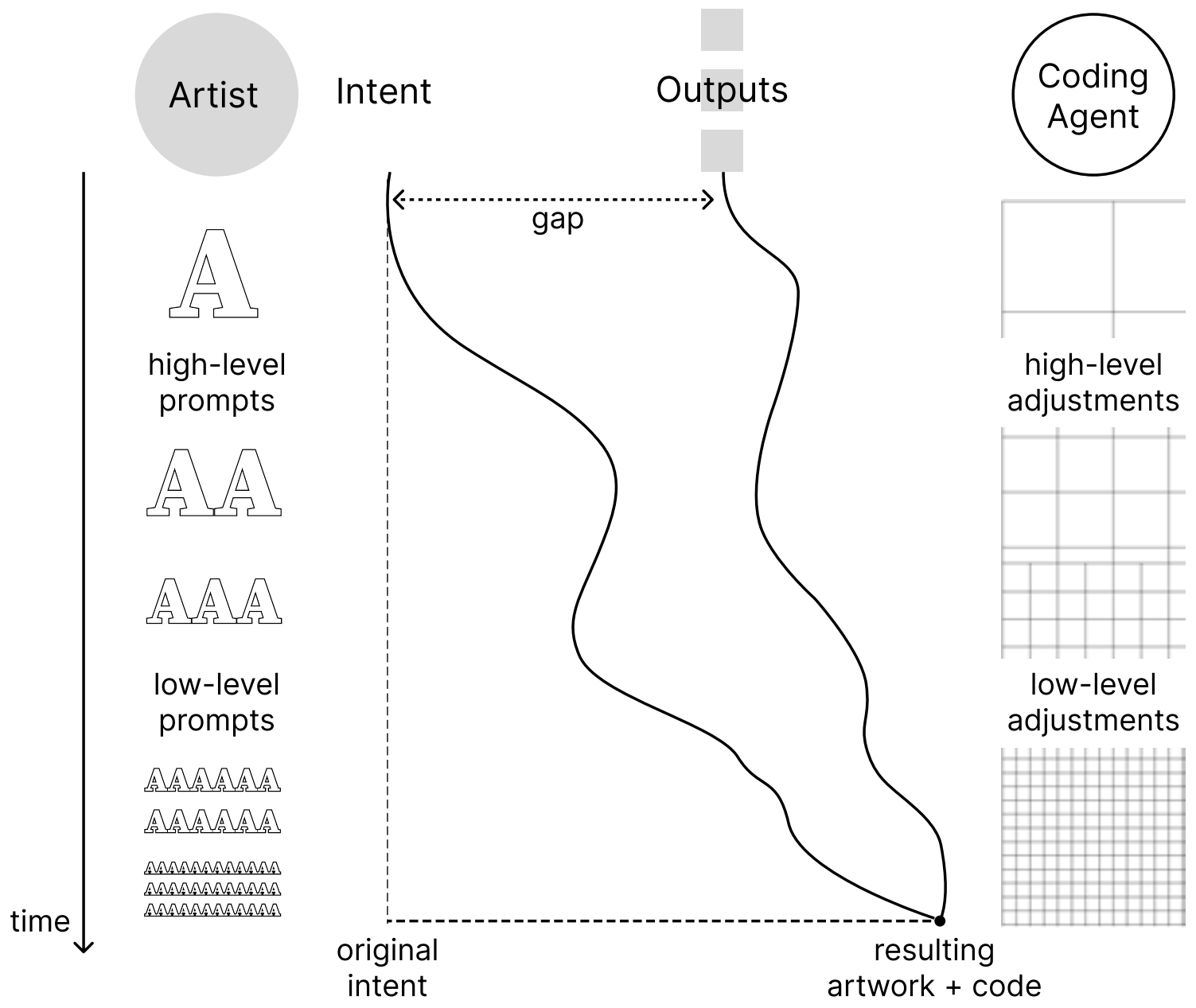

**Fig. 4** - Involving a coding agent could skew the artist's original intent.

## 3 Methodology and approach

Based on these reflections on the nature of generative art and its combination with coding agents, I've started exploring the main research question as an artist and researcher:

*How does the use of LLM-based coding agents change the practice of code-based generative art, compared to a purely manual approach?*

In my investigation I've narrowed it down to two aspects:

**RQ1) How does the use of coding agents influence the time required and the nature of outputs?**

**RQ2) How does it interact with a generative artist's own process and technical skills?**

As overarching study structure, I decided on an evocative autoethnographic approach (Bochner and Ellis 2016), a methodology for illuminating broader cultural experiences through representing lived ones. Given my own status as a practicing generative artist and researcher, I was in the unique position to provide an artist's interior perspective on the impact of this technology on my work. Using creative expression as both a medium for reflection and a unique subjective representation fit well: previous studies on reflective approaches through visual practice (Jung and Trischler 2021) or design memoirs (Devendorf et al. 2020) show the value of such evocative autoethnography, also for difficult and potentially threatening scenarios.

[7] https://antigravity.google

Given the specific nature of generative art, I also expected to gain more insights from having direct access to a practitioner's internal point-of-view. Having experience with the qualities of the material (Malafouris' clay (Malafouris 2008) if you will) would give me better access to the changes taking place.

## 3.1 Study setup and personal background

I used the *#Genuary 2026* art challenge to perform this research: during *#Genuary*, which happens every January, artists are expected to post a new generative/creative coding artwork every one of the month's 31 days based on a specific artistic prompt and share the results on social media. The prompts range from straightforward and/or technical (Day 10: "Polar coordinates"), to ambiguous and open to interpretation (Day 14: "Everything fits perfectly", Day 19: "16x16").

The daily format comes with its own constraints: creating a generative artwork means taking the time required out of one's schedule and producing something new every single day.

I had some experience when it comes to daily challenges and *#Genuary* in specific. I had participated in three *#Genuarys* before (2022, 2023 and 2025) and had kept up a practice of posting a new generative artwork on social media every day for almost two years over the course of 2022 and 2023[8]. Therefore, I had experience with quickly producing generative artworks and an existing setup based on web technologies like HTML, CSS and JavaScript.

As for including coding agents into my artistic practice: I had performed an informal experiment during summer 2025 with OpenAI's ChatGPT-4. While simple requests worked, asking for more complex algorithms and systems would – even on first try - lead to broken code. And some internal limit of complexity would make the system no longer being able to add features or fix bugs after a certain number of lines of code had been reached which would make me decide against continued use.

During this study I would have access to various coding agents and LLMs:

- OpenAI ChatGPT-5.2 (Thinking) via OpenAI's website[9].
- Google Gemini Pro 3 (High) in Google's Antigravity IDE.
- Anthropic's Claude Opus 4.5 (Thinking) in either Google's Antigravity IDE or in the Claude Code expansion for the VSCode IDE.

I'd be free to access whichever agent I prefer. Since providers apply a quota for each model, I would also be able to switch between them when I exceeded it.

My process of coding a generative artwork would follow roughly the approach described in section 2.1.: trying to get to something visible as quickly as possible, then shrinking the gap between intent and output by adjusting the code on an increasingly finer-grained level. Once satisfied with the result, I would capture a video of it and post it to social media.

---

[8] All visible on my Instagram account https://instagram.com/dominikus
[9] https://chatgpt.com

## 3.2 Methodology

The 31-day study was split up into three phases:

- **Phase 1 (days 1 - 4)**: For the first days I participated in *#Genuary* without a coding agent. This phase would enable me to better evaluate the difference between the two conditions.
- **Phase 2 (days 5 - 15)**: The next eleven days let me get used to a coding agent during *#Genuary*. I freely used it and structured my interaction and approach to the challenge however I wanted. In parallel, I developed a structured approach for the last phase.
- **Phase 3 (days 16 - 31)**: In the final part of the challenge I switched to the structured approach that had been developed during Phase 2: I started each day's session with a sketch on paper of what I envisioned. I then started prompting the coding agent. At any point, I could also change the resulting code. I kept track of duration and wrote notes of emotions/thoughts throughout. Processing a prompt took the coding agent around 30 seconds each, which gave me ample time to do that.

My approach resulted in the following data and artifacts for each of the 31 days:

- Resulting artwork and code
- Initial sketch (from day 16 onwards)
- Prompt history and responses (from day 5 onwards)
- Duration
- Notes on thoughts and emotions (from day 16 onwards)
- Notes on existing algorithms/techniques used for a given piece

I analyzed the notes first with an initial coding approach to turn the qualitative data into codes, then with focused coding to extract the most salient aspects and categorize them (Saldaña 2013) and identified three main themes that are discussed in section 5.

# 4 Results

I ended up spending a total amount of roughly 46 hours working on the prompts (an average of 1:30 hours on each challenge). The challenge resulted in 12,983 lines of code (outside of libraries used) and 10,832 prompted words from myself.

## 4.1 General observations

What I had noticed in my informal experiment from last summer was no longer the case: even complex requests practically always led to working code that would fulfill what had been asked for. Also, the limit of complexity mentioned above was no longer there: even more complex ideas that would combine multiple algorithms could be extended and worked on for longer.

The resulting code usually followed a sensible structure and made customizing it relatively easy. One aspect, for example, was that overarching parameters such as color palettes and simulation parameters would be defined at the top of the file, thus making experiments and tweaking easy. Also, coding agents tend to add many comments (Fawareh et al. 2024) which again helps with understanding the resulting code. Therefore, the results let me switch from prompting to changing the code whenever I wanted. This was also supported by the setup of the coding agents themselves, that would (usually) realize when I had manually changed things and respect those changes.

After some period of initial experimentation, I would prefer the Claude Opus 4.5 model for my work and use it until I ran into quota limitations, falling back to Gemini Pro 3. I found that Opus gave the best results for the aspects mentioned above.

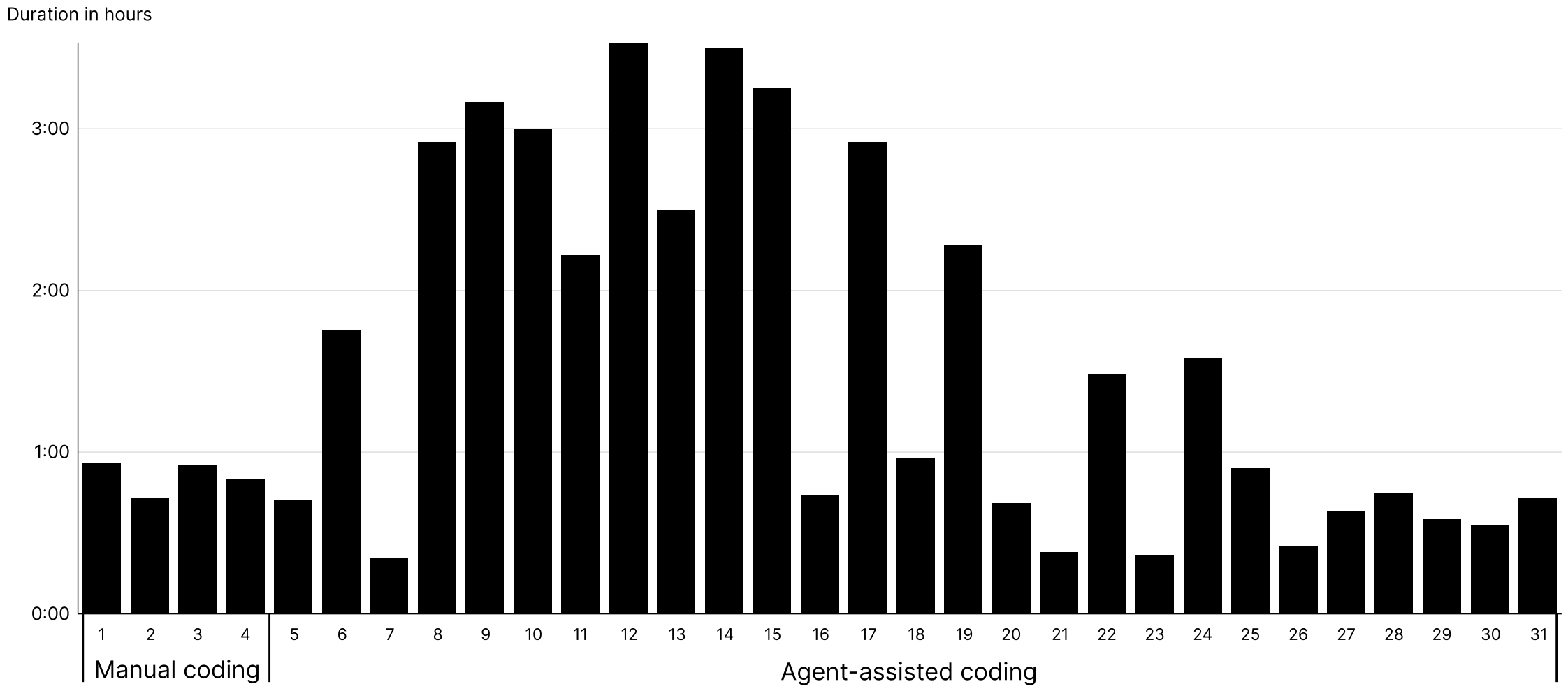

**Fig. 5** - Hours spent on each daily artwork.

## 4.2 Changes in the practice

### Time required and nature of outputs

Regarding the first research question, one assumption in using a coding agent might be that it just speeds things up. Which I found to be the case. However, in addition, it enables a much freer exploration of ideas which would then extend the time required again.

As seen in **Fig. 5**, while my manual coding hovered around one hour each day, my initial uptake of coding agents would extend this time greatly, sometimes to more than three hours. Looking back at the notes and logs, this duration came in a few instances from frustrating bugs and not being able to reach the original intent, but most of the time was spent on pure experimentation: pushing the code and the agent into different directions and exploring various existing algorithms and techniques (see **Table 1**).

**Table 1** - Named techniques used during the 31 days.

| Day | Named techniques used | Day | Named techniques used | Day | Named techniques used |
|---|---|---|---|---|---|
| 1 | | 12 | Collisions (matter.js) | 23 | Noise, Chaikin curves |
| 2 | Collisions (matter.js) | 13 | Camera, particles | 24 | Path tracing |
| 3 | | 14 | Path tracing, Voronoi | 25 | L-Systems |
| 4 | Noise, Flow field | 15 | Path tracing | 26 | |
| 5 | Collisions (matter.js) | 16 | Reaction-diffusion | 27 | Cellular automata |
| 6 | | 17 | Noise, Flow field | 28 | Noise |
| 7 | Noise, Flow field | 18 | Quadtree Subdivision, Agents | 29 | Cellular automata |
| 8 | Path tracing | 19 | Flow field | 30 | Quadtree Subdivision |
| 9 | 3D cellular automata | 20 | Chromatic aberration | 31 | Path tracing |
| 10 | Noise, Hyperbolic coordinates | 21 | Quadtree Subdivision | | |
| 11 | Particles | 22 | | | |

All days had me follow through with my original idea/sketch (see **Fig. 6**) except Day 22 (Pen-plotter ready): I had started with a sketch inspired by Ruth Asawa's looped-wire sculptures[10] and spent over one hour before giving up and switching to an idea that proved to be more attainable for the coding agent. While the agent kept producing working code, the results were simple geometric line drawings - far away from my sketch. Even uploading an image of one of Asawa's sculptures at one point couldn't have it succeed – an obvious change to my artistic intent caused by the tool.

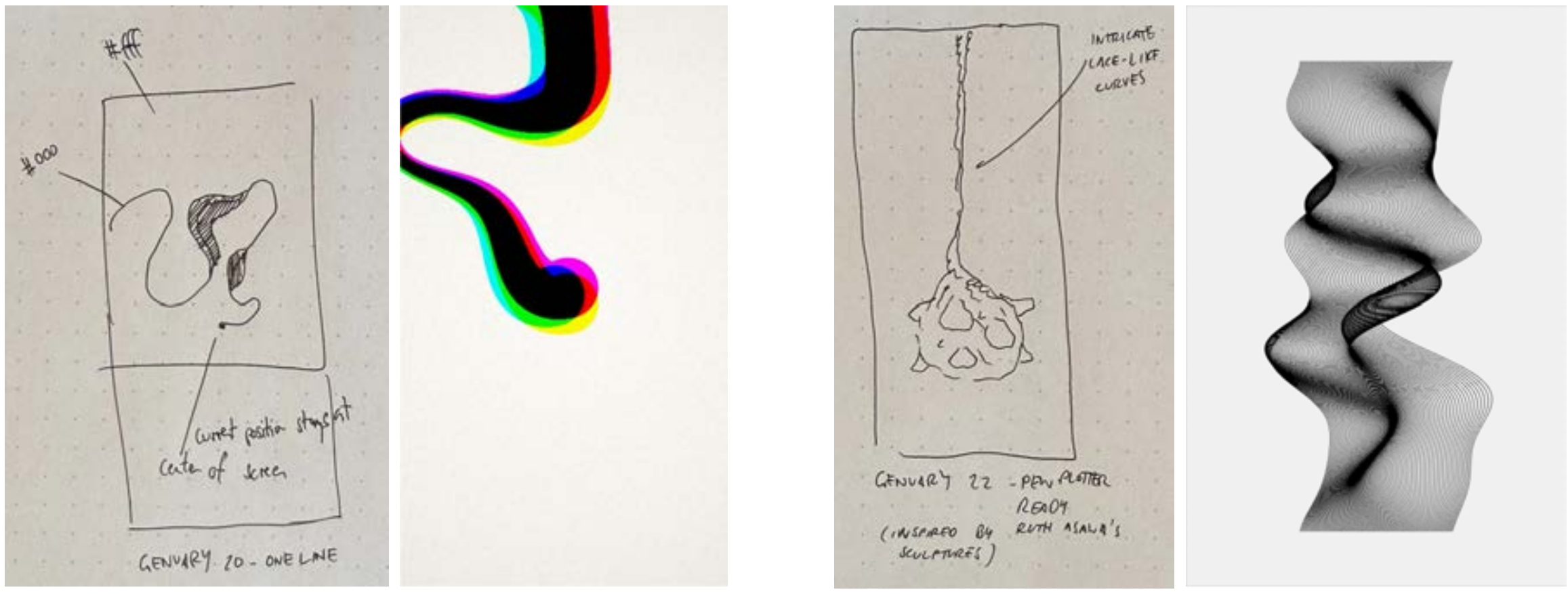

#Genuary 20: One line. #Genuary 22: Pen-plotter ready

**Fig. 6** – Original sketches and resulting artworks. Note that the left artwork is originally animated.

Something that often happens in generative art due to its reliance on emergence and surprises is that an envisioned system yields different outputs than expected. This was something that I also regularly encountered during the 31 days: my sketch and its proposed algorithm would lead to not nearly as interesting results as I had hoped for in the actual implementation. Since time is limited within such daily challenges, this would - in manual coding - often lead to increasingly desperate parameter tweaking, simply because larger changes are no longer feasible. With coding assistance, however, I could ask for bigger changes (like the introduction of an additional algorithmic technique such as a noise field, for example) and have the machine quickly add them to the existing system.

**Interaction with process and technical skills**

My own process on a more granular level changed from pure coding to switching between prompting and coding (see **Fig. 2** (right)). Which of the two I chose was almost always a matter of the expected efficiency for a prompt: much of the initial setup for each project was writing boilerplate code to set up the various parts required. Writing a prompt of a few dozen words at this point would thus produce hundreds or even thousands of commands in code. This made up the main source of gains that using the coding agent brought. Most days, I would start with a prompt like this:

> *"please generate a stand-alone html file that shows a 9:16 canvas with 20px margin. run a reaction-diffusion simulation in a fragment shader and display the results on the canvas (background dark, foreground white-ish). don't use any libraries. also generate an irregular grid that's overlain on the reaction-diffusion simulation. every cell of the grid uses different rules/parameters for the reaction-diffusion simulation."* (Initial, unaltered prompt for Day 16, "Order and disorder")

[10] https://ruthasawa.com/art/sculpture/

This would usually produce hundreds of lines of HTML-scaffolding, JavaScript- and fragment shader code from this single short prompt. During the process, however, this "transmission ratio" between prompt and resulting code would gradually worsen, up to a point where changing things directly in the code (e.g., a single color in a color palette) would be faster than trying to explain that to the agent.

Nevertheless, I noticed that over the course of the 31 days the interaction with the coding agent would gradually nudge me towards prompting more and giving vaguer instructions:

On Day 21 I would ask for a "*Bauhaus-inspired palette*", something which might have been done through a Google Search in a manual setup but would in any case have required more thought. Now, this creative decision of which colors to pick was left completely to the agent.

Even more extensive, on Day 29 I would tell the coding agent: "*the lifeforms themselves should have a blobby, meandering look (maybe use metaballs)… give everything an oldschool, sepia biology book from the 1800s*" and simply accept its ideas for titles, an additional frame around the piece, etc.

Nevertheless, technical skills would still prove useful in events when the coding agent would get stuck on certain bugs. Having an idea about the actual algorithm and common implementation details would often help me in describing potential error sources to the coding agent and thus arrive faster at a solution or free it from some misguided bug-tracking loop. Often, I would also fix the problem myself directly in the code, without any agentic involvement.

In certain occurrences the coding agent would suggest that a problem did not really exist – for example, instead of addressing broken rendering logic it would imply that the contrast between foreground/background was too low to see anything etc. Having a clearer understanding of what's happening and being able to use browser developer tools for exploration helped me here in getting the coding agent to fix bugs.

# 5 Discussion

My personal observations over the course of the 31 days feed into a deeper reflection on how the use of coding agents might change the practice of code-based generative art and hint at a potential degradation: an initial trend towards wide but shallow results which shapes the practice into parametric instead of code-based art and might finally lead to an overall artistic surrender.

## 5.1 Wide but shallow

The use of a coding agent and the difficulty collapse it brings enables much more ambitious projects in the very restricted amount of time available for a single *#Genuary* project: prompting the machine with "a jumble of variously colored lines in motion" is as complex as prompting it with "a reaction diffusion-simulation on the fragment shader" while the outcomes of both prompts vary extremely in levels of complexity.

This was what I originally had hoped for: a tool that would take off the strain and frustrations that sometimes arise in coding and let me freely explore whatever ideas would pop into my head. While before I would have applied mental blinders cutting off certain ideas in the knowledge that I would have to produce the code myself, coding assistance enabled a more free-flowing idea generation. The initial experimental phase indeed felt like entering a completely open field: all the complex algorithms I had seen producing interesting visual results but seemed too daunting to implement[11] were only one prompt away. All art practices – including code-based art - require a certain expertise to reach mastery, yet here it felt like such "mastery" only required knowing the right magic words to type into the machine.

This aspect frees artists from the necessity to excel at coding when attempting to use complex algorithms in their bricolage and opens up even the most complex ones to everybody. Potentially, this would even enable people who have never learned to program to create code-based art.

Yet it also hints at the downside of this widening of possibilities: manually implementing a complex algorithm requires understanding all its parts intricately. Due to emergent properties of the underlying computation, small changes in its implementation can lead to vast changes in outcome. Therefore, to be able to influence and shape such an algorithm's output to one's intent, artists must be knowledgeable in these intricacies – something that feels no longer necessary with a coding agent.

Therefore, using coding agents in generative art inevitably leads to wide but shallow results: the coding agent's tendency towards statistical averages produces the "normal" version of the algorithm and the further away from this average the requested results, the harder it becomes. So suddenly every artist can work with - for example - reaction-diffusion simulations, yet the results will be all more or less the same. Without understanding and mastering an algorithm it cannot properly be used for individual expression, just like one must go through the tedious and complex task of learning and understanding an instrument to be able to play unique music. Skipping the learning part leaves one stuck with pressing Play on a record player.

I noticed this myself in the study: as depicted in **Table 1** I would regularly fall back on path tracing (5 times) or cellular automata (3 times). Also, my time spent experimenting would significantly shorten from day 20 onwards (see **Fig. 5**) from both a certain exhaustion with the challenge but also from the ability to generate interesting outputs from barely understood algorithms.

The pull towards broad, uninformed use of algorithms also leads to a more uninformed treatment of the central aspect of code-based generative art: the code itself.

## 5.2 Parametric instead of code-based art

What I noticed in my own practice was that looking at and editing the code would happen later and later in the process. As I outlined above, the "transmission ratio" from prompts to code would have to be so low as to make such manual changes worthwhile.

Additionally, my prompts would become increasingly vague, leaving larger artistic choices to the machine. While the question whether a computer can be an artist at all has been around since the beginnings of generative art (cf. McCormack et al. 2014), coding agents shift the quality of the computer's role in generative art substantially. The question depicted in **Fig. 4** whether the process skews an artist's intent is now more about if an intent existed at all.

Editing the code only towards the end changes code-based art to parametric art: existing algorithms are used for the setup of the structure, larger changes are implemented by the agent (with the artist not necessarily being aware of how it works) with some manual fine tuning of more aesthetic parameters (color palettes, line widths, etc.) towards the end. The descent along the ladder of abstraction towards more detailed adjustments (cf. **Fig. 3**) is pre-empted and artists only briefly hop onto the lowest rung. Aspects of code improvements such as code golf also no longer fit in with this approach to the practice.

---

[11] Spending a whole weekend on writing a ray tracer (Shirley et al. 2025), for example, seems more like a threat than a promise.

Prompting also puts an end to inventing novel algorithms. While much of generative art is built on the bricolage idea where existing algorithms are re-used, it is also common to invent approaches from scratch (e.g., drawing lines or circles based on specific rules that one comes up with). As generative artists know, however, this approach often results in visually uninteresting outputs, since emergence is very hard to predict. Additionally, specifying such novel ideas cannot be made much shorter in a prompt than directly specifying it in code anyway. Thus, a turn towards coding agents automatically discourages coming up with novel algorithms. Instead, it seems much more effective to learn the names of the various existing algorithms (the "magic words") and have the machine reliably produce visually interesting (though repetitive) outputs.

One could well imagine a future "generative art" tool that presents a list of existing algorithms to mix-and-match from and finally lets the "artist" drag a few sliders to adjust the parameters. This abandonment of the code could ultimately lead to an even more substantial change:

### 5.3 Artistic surrender

Along the lines of Shaw's and Nave's argument for "cognitive surrender" where people are "adopting AI outputs with minimal scrutiny, overriding intuition and deliberation" (Shaw and Nave 2026) this might lead to an overall "artistic surrender" where coding agents are not only used for producing code but also slowly take over more and more artistic decisions.

Since coding agents are geared towards a mainstream audience, they do not perform the repetitive requests for clarification that previous generations of chatbots exhibited before giving an answer. Instead, they are producing outputs from the first prompt onwards, taking decisions in areas that have not been specified. Thus, an artist might not even realize which aspects of the project have been decided by the machine thus surrendering parts of their practice without any awareness of it. This induced laziness might in the long term lead to a decrease in overall skill (Shen and Tamkin 2026) and shrink variety within the whole field, since understanding algorithms is required to find unique expressions, as outlined above.

This threatens generative art as a specific art form with it being at its core about emergence and surprises through computation. Discouraging the in-depth interaction with the code fundamentally changes the practice. To build on Malafouris' pottery metaphor mentioned above: the quality of the code material is changed drastically since artists interact with the coding agent's material instead. Working as a generative artist with a coding agent is more akin to industrial pottery than sitting at a potter's wheel. The results might look the same to an outsider, but the internal flow and the "grain" of the process are completely different.

## 6 Conclusion

In this paper, I presented an autoethnographic study on the effect of using coding agents on my generative art practice. I identified the concepts of difficulty collapse as an enabler for wider but ultimately shallower artworks as well as the pull towards a more parametric than code-based practice and ultimately a potential for full artistic surrender.

Future work could see the limitations of this study (it being mainly an in-depth look at one artist's practice) addressed, by applying similar methods to more diverse sets of participants with distinct experience levels, skill sets, approaches to the art or in a more longitudinal approach. Similarly, the development of artistic tools and user interfaces based on the results of this study that might mitigate the identified risks would be another fruitful research direction.

Realistically, coding agents have already or will soon become part of every coder's practice, artist or not – the benefits simply cannot be ignored (e.g. Cui et al. 2025). Yet one must be aware of the potential downsides. For artists, more experimentation and coming up with interesting ideas by oneself should stay the default. Having a coding agent produce an algorithm can be a convenient shortcut but should be treated like copying the code from the internet: as a first step to further experimentation based on in-depth understanding to keep the domain of personal expression that is generative art from being replaced with the bland agreeableness of large-language models.